\documentclass[conference, a4paper]{IEEEtran}
\IEEEoverridecommandlockouts
\usepackage{cite}
\usepackage{amsmath,amssymb,amsfonts}
\usepackage{algorithmic}
\usepackage{graphicx}
\usepackage{textcomp}
\usepackage{xcolor}

\usepackage{multirow}
\usepackage{booktabs}
\usepackage{url}
\usepackage{hyperref}
\usepackage{subfigure}
\usepackage{svg}

\def\BibTeX{{\rm B\kern-.05em{\sc i\kern-.025em b}\kern-.08em
    T\kern-.1667em\lower.7ex\hbox{E}\kern-.125emX}}
\begin{document}

\title{Adaptive Hybrid Spatial-Temporal Graph Neural Network for Cellular Traffic Prediction\\
}

\author{

\IEEEauthorblockN{Xing Wang, Kexin Yang, Zhendong Wang, Junlan Feng, Lin Zhu,
Juan Zhao, Chao Deng}
\IEEEauthorblockA{JIUTIAN Team, China Mobile Research Institute, Beijing, China}
\IEEEauthorblockA{
Email: \{wangxing, yangkexin, wangzhendongai, fengjunlan, zhulinyj, zhaojuan, dengchao\}@chinamobile.com}
}


\maketitle

\begin{abstract}
Cellular traffic prediction is an indispensable part for intelligent telecommunication networks. Nevertheless, due to the frequent user mobility and complex network scheduling mechanisms, cellular traffic often inherits complicated spatial-temporal patterns, making the prediction incredibly challenging. Although recent advanced algorithms such as graph-based prediction approaches have been proposed, they frequently model spatial dependencies based on static or dynamic graphs and neglect the coexisting multiple spatial correlations induced by traffic generation. Meanwhile, some works lack the consideration of the diverse cellular traffic patterns, result in suboptimal prediction results. In this paper, we propose a novel deep learning network architecture, Adaptive Hybrid Spatial-Temporal Graph Neural Network (AHSTGNN), to tackle the cellular traffic prediction problem. First, we apply adaptive hybrid graph learning to learn the compound spatial correlations among cell towers. Second, we implement a Temporal Convolution Module with multi-periodic temporal data input to capture the nonlinear temporal dependencies. In addition, we introduce an extra Spatial-Temporal Adaptive Module to conquer the heterogeneity lying in cell towers. Our experiments on two real-world cellular traffic datasets show AHSTGNN outperforms the state-of-the-art by a significant margin, illustrating the superior scalability of our method for spatial-temporal cellular traffic prediction.
\end{abstract}


\section{Introduction}
Total global mobile data traffic reached 67EB per month by the end of 2021, and is projected to grow 4.2 percent to reach 282EB per month in 2027 \cite{ericsson}. 
The explosive growth of traffic not only brings a huge demand for network capacity, but also brings challenges for telecom network management and resource allocation. As a crucial aspect of telecom network operation, traffic prediction is essential for intelligent wireless networks. Accurate cellular traffic prediction plays an important role in network planning, traffic scheduling, network fault diagnosis and reducing operation cost, etc. 


\begin{figure}[htpb]
\begin{center}
\includegraphics[width=7cm]{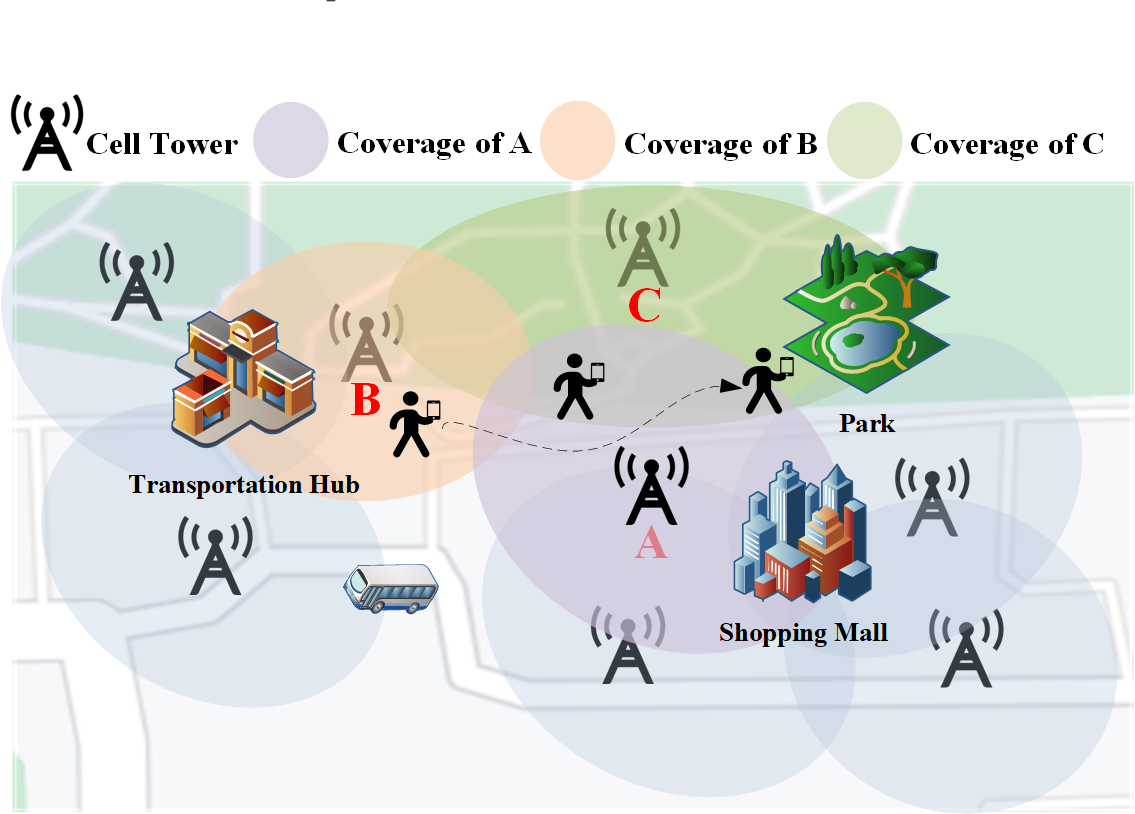}
\end{center}
\caption{An example of cell tower distribution in a region. Purple, orange and green ellipses indicate the coverage of the cell tower A, B and C. Blue ellipses means the coverage of other cell towers. The coverage areas of different cell towers are partially overlapped, so users can switch between cell towers and still maintain the connection.}
\label{Coverage}
\end{figure}


However, it is extremely challenging to predict cellular traffic due to several reasons. 
First, cellular traffic exhibits nonlinear temporal dependencies since the mobile data traffic is extremely dynamic. For instance, a user can consume a large volume of data for a moment via a given cell tower. At the next moment, this user may stop the connection or migrate to a new cell tower, causing a certain amount of traffic to disappear suddenly \cite{Wang2021adaptive}. The discontinuous nature of data usage makes the inherent temporal dependency of mobile traffic a complex nonlinear and unstable problem. 
Second, frequent user mobility and complicated network scheduling mechanisms bring complex spatial correlations between cell towers. As shown in Fig.  \ref{Coverage}, different cell towers in a wireless network maintain distinct coverage zones. Users can attach to a cell tower within its coverage area to access mobile network services and consume traffic. Due to the limited coverage of wireless signals, users could switch between multiple cell towers as they migrate between areas, resulting in the spatial correlations of cellular traffic. Moreover, users can easily travel across the city within half an hour with the efficient urban transportation, which brings the spatial dependencies even with distant cell towers \cite{wang2018spatio}. For better user experience, the wireless network scheduling mechanism may also hand over the user to a closer cell tower, or a cell tower with fewer users, or a distant cell tower with stronger signal, etc., which increases the complexity of spatial correlation.
Third, 
the different capacities, geographical locations, and surroundings of cell towers make their data traffic patterns diverse, which called heterogeneity. For example, the data traffic of a cell tower near a shopping mall shows significant increase at weekend compared with the weekday. Meanwhile, it is opposite for the cell tower located in a subway station, whose data traffic in the weekday is regularly higher than that of the weekend with obvious morning and evening peak. 

Recently deep learning-based methods for mobile traffic prediction have attracted the interests of researchers due to its powerful ability to capture intricate data patterns. Typically, Recurrent Neural Networks (RNN) based models, such as Long-Short Term Memory (LSTM) networks \cite{6795963}, have shown promising performance on modeling temporal dependencies \cite{9148738}. But they ignore spatial correlations, resulting in a loss of accuracy. 
\cite{8667446} proposed the use of convolutional LSTM (ConvLSTM) \cite{shi2015convolutional} for spatial-temporal traffic prediction, which adopts a grid-based assumption that cells have a Euclidean spatial relationship. 
Although the grid-based approach achieved better performance than LSTM, it may not accurately capture the spatial dependencies of the cell towers.
Several research tend to employ Graph Neural Networks (GNNs) \cite{ZHOU202057} for modeling the spatial dependencies based on graph. \cite{Wang2021adaptive} proposed to implement graph convolutional networks (GCN) \cite{kipf2017semi} with a static predefined graph constructed based on the geographical distance between base stations, to capture the spatial correlations of cellular traffic. \cite{9625773} adopted an attention mechanism to capture global spatial-temporal correlations for the dynamic characteristics of cellular traffic. While the remarkable results have been achieved, they still have limitations. The static graph can hardly reflect the time-varying relationships between nodes. And the dynamic graph generated approaches often have high model complexity and rely heavily on the traffic data, which is sensitive to data noise and hard to converge. \cite{Wang2021adaptive} implemented a spatial-temporal joint convolutional network with adaptive multi-receptive fields for capturing the heterogeneity in mobile traffic. But it only focus on modeling node-specific patterns of the spatial feature propagation and ignore the nodes' distinct spatial-temporal perceptions.

In this paper, we propose a novel Adaptive Hybrid Spatial-Temporal Graph Neural Network (AHSTGNN) to address the aforementioned challenges. 
We apply an Adaptive Hybrid Graph Learning Module (AHGLM) to 
capture the complex spatial dependencies in
the cellular network motivated by the observed daily routine of user behavior. The majority of users regularly go to several locations, such as their workplaces or homes, which implies relatively static spatial dependencies between their visited cell towers. Meanwhile, they may spontaneously visit other locations, indicating the occasional dynamic spatial relationships coexisted. 
We combine the Static Adaptive Graph Learning (SAGL) and Dynamic Graph Learning (DGL) to capture the stable and occasional relationships in an AHGLM, respectively.
Meanwhile, we implement a Temporal Convolution Module (TCM) equipped with multiple periodic components to enhance the model ability to capture the non-linear temporal features. For solving the heterogeneity, we construct a Spatial-Temporal Adaptive Module (STAM) to model the various spatial-temporal perceptions of cell towers. 
Overall, our main contributions are summarized as follows:
\begin{itemize}
\item We design a novel AHGLM to capture the compound spatial dependencies and a TCM with multi-periodic data inputs to model the nonlinear temporal dependencies in cellular traffic.
\item We propose an effective framework AHSTGNN for accurate cellular traffic prediction with the STAM for modeling the various traffic patterns, which can effectively protect the nodes from unexpected spatial noise.

\item Our experiments on two real-world cellular traffic datasets show AHSTGNN outperforms the state-of-the-art, illustrating the superior scalability of our method.
\end{itemize}


\section{DATASET DESCRIPTION and PRELIMINARIES}
\subsection{Dataset Description}
The Jiangsu and the Milan \cite{barlacchi2015multi} are two cellular traffic datasets analyzed in this paper. The Jiangsu dataset is collected by a Chinese mobile network operator, China Mobile, from January to March, 2021. The data is collected from 1,051 cell towers in Jiangsu Province, China, which contains the cellular traffic data with sampling rate per 15 minutes. 
The Milan dataset is one of the most commonly used public datasets in cellular traffic prediction. This dataset provided by Telecom Italia is recorded from November 1, 2013 to January 1, 2014 with a 10-minute time slot. It contains the cellular traffic collected from the $100 \times 100$ cells subdivided in Milan, where the size of each cell is $235m \times 235m$.

\subsection{Preliminaries}
In this paper, we formulate the cellular traffic prediction problem as a spatial-temporal sequence prediction problem. Specifically, we define a graph $\mathcal{G}=(V,E,A)$ as the traffic network. $V$ denotes the set of cell towers, where $|V|=N$ ($N$ indicates the number of vertices). We can also call them nodes in the rest of this paper. $E$ is the set of edges which represents the spatial proximity between nodes. $A\in\mathbb{R}^{N\times N}$ is the adjacency matrix of $\mathcal{G}$, where $A_{v_i,v_j}$ represents the connection between node $v_i$ and $v_j$. The data traffic of the cell towers, also known as graph signal matrix, is denoted as $X_t=(x_t^1,x_t^2,\cdots,x_t^N)^T \in \mathbb{R}^{N\times F}$, where $F$ indicates the feature dimension and $t$ denotes the time step, $x_t^i$ represents the observations of the $i$-th cell tower at time step $t$.

\textbf{Problem Defined.} Given the graph signal matrix of historical $\tau$ time steps $X=(X_{t-\tau+1}, \ldots,X_{t-1}, X_{t})\in \mathbb{R}^{\tau \times N \times F}$, our target is to predict the future data traffic for all $v \in V$ in the next $M$ time steps $\hat{Y}=\left(\hat{X}_{t+1},\hat{X}_{t+2},\ldots,\hat{X}_{t+M}\right)\in \mathbb{R}^{M\times N\times F}$. In general, we formulate the problem as learning a mapping function $\mathcal{F}$ to  map the graph signal matrix of historical time steps to the graph signal matrix of future time steps:
\begin{equation}
    \left(\hat{X}_{t+1},\hat{X}_{t+2},\ldots,\hat{X}_{t+M}\right)=\mathcal{F}_{\theta}\left(X_{t},X_{t-1},\ldots,X_{t-\tau+1};\mathcal{G}\right),
\end{equation}
where $\theta$ denotes all the learnable parameters in our model.

\section{METHODOLOGY}
\subsection{Architecture}
\begin{figure}[htpb]
\begin{center}
\includegraphics[width=8.5cm]{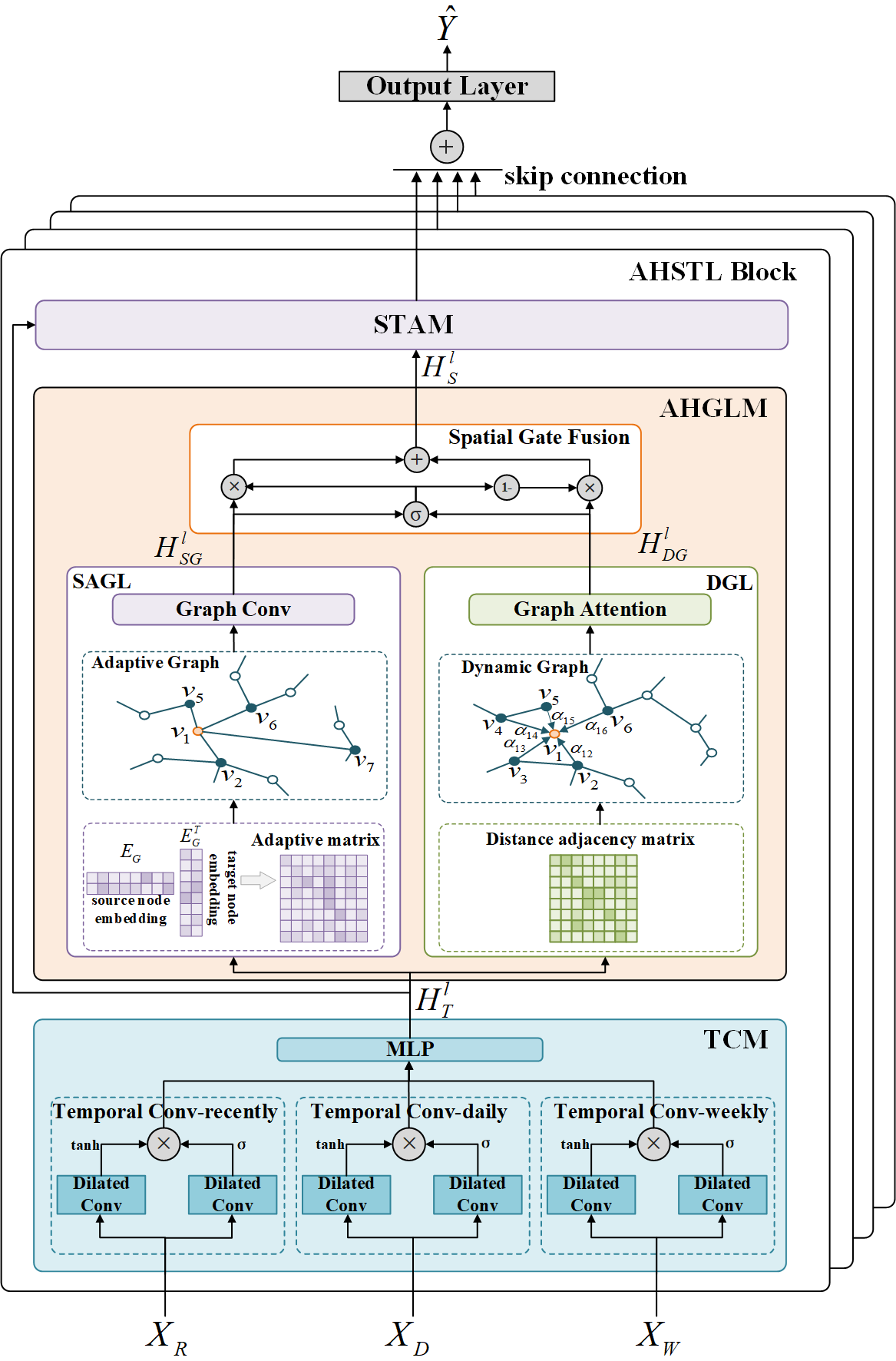}
\end{center}
\caption{The Architecture of AHSTGNN.}
\label{AHSTGNN}
\end{figure}

Fig.  \ref{AHSTGNN} illustrates the architecture of our AHSTGNN model, which consists of multiple Adaptive Hybrid Spatial-Temporal Learning Blocks (AHSTL block) with skip connections, and an output layer. Each AHSTL block is composed of a Temporal Convolution Module (TCM), an Adaptive Hybrid Graph Learning Module (AHGLM), and a Spatial-Temporal Adaptive Module (STAM). Among them, the TCM with multi-periodic data inputs is used to extract different periodic characteristics of the data traffic in the temporal axis. The AHGLM is adopted to capture the complex spatial dependencies between cell towers. We also incorporate the STAM to adaptively model the spatial-temporal tendency of cell towers for solving the heterogeneity problem.

\subsection{Periodic Input Data}
We introduce three periodic inputs for capturing multi-period temporal dependencies. $T_R$, $T_D$ and $T_W$ represent the input length of  historical time steps for the recently, daily and weekly components in TCM respectively. $L_D$ and $L_W$ denote the number of past days and past weeks we select, respectively.

Intuitively, future mobile data traffic are influenced by the most recent traffic sequence trend. Assume $t$ is the current time step and $q$ is the sampling frequency per day. The recently component extracts the features of cellular traffic adjacent to the predicting period. 
$X_R=\left[X_{t-T_R+1},X_{t-T_R+2},\ldots,X_{t}\right]\in \mathbb{R}^{T_R\times N\times F}$ denotes the recent data traffic input.
The daily component models the daily patterns of cellular traffic in the same period as the predicting period over the last few days.
For the ${L_d}^{th}$ day, we denote 
$X_d=\left[X_{t-L_d\times q+1},X_{t-L_d\times q+2},\ldots,X_{t-L_d\times q+T_D}\right]\in \mathbb{R}^{T_D\times N\times F}$ as the daily traffic input, where ${L_d} \in\{1, \cdots, {L_D}\}$. For the last ${L_D}$ days, we simply sum all the corresponding ${X_d}$ to get ${X_D}$.
The weekly component extracts the periodic features of cellular traffic in last few weeks with the same weekday and same time period as the future prediction time period. For the ${L_w}^{th}$ week, we also have 
$X_w=\left[X_{t-7\times L_w\times q+1},X_{t-7\times L_w\times q+2},\ldots,X_{t-7\times L_w\times q+T_W}\right]\in \mathbb{R}^{T_W\times N\times F}$, where ${L_w} \in\{1, \cdots, {L_W}\}$. To obtain the weekly features of the last ${L_W}$ weeks, we use the same way as ${X_D}$ to get the ${X_W}$ as the input of the weekly component.

\subsection{Adaptive Hybrid Spatial-Temporal Learning Block}

As shown in Fig. \ref{AHSTGNN}, 
we denote the input of the $l^{th}$ AHSTL block as $H^{l-1}$, including three periodic data features, the recently feature $H_{R}^{l-1}$, the daily feature $H_{D}^{l-1}$ and the weekly feature $H_{W}^{l-1}$, which are the outputs of the TCM in the $(l-1)^{th}$ block. The outputs of TCM and AHGLM in the $l^{th}$ block are represented as $H_{T}^l$ and $H_{S}^l$. After the STAM, we obtain the output of the $l^{th}$ block, denoted as $H^l$. For consistency, we denote the hidden state feature dimension as $D$ and the historical time steps of data traffic as $T$. 

\subsubsection{Temporal Convolution Module}
Temporal Convolution Module consists of three gated Temporal Convolution Networks (gated TCN) \cite{dauphin2017language}, named Temporal Conv-recently, Temporal Conv-daily, and Temporal Conv-weekly, to capture the complicated temporal correlations and periodicity of the data traffic. 
 The initial input for the $1^{th}$ TCM are the periodic input data $X_R, X_D, X_W$ described above.
The temporal convolution is highly parallelizable and amenable to maintaining higher computational efficiency. By stacking multiple layers, TCN allows for an exponential receptive field with dilated causal convolution. The gating mechanism shows the capability of handling sequence data and enhances the modelling capacity of temporal convolution. Taking the recently component as an example, it takes the form:
\begin{equation}
H_R^l=tanh\left({W}_{t1}* H_R^{l-1}+ {b}_{t1}\right) \odot \sigma\left({W}_{t2} * H_R^{l-1}+{b}_{t2}\right),
\end{equation}
where $*$ denotes the convolution operator, ${W}_{t1}$, ${W}_{t2}$, ${b}_{t1}$ and ${b}_{t2}$ are learnable parameters, $\odot$ is element-wise product, $\sigma$ means sigmoid activation function and $tanh$ means hyperbolic tangent activation function. The daily and weekly component share the same framework as the recently component. Afterwards, the temporal features are concatenated into one vector, then fed into a Multilayer Perceptron (MLP). The final output of the $l^{th}$ TCM is $H_T^l\in\mathbb{R}^{T\times N\times D}$. 

\subsubsection{Adaptive Hybrid Graph Learning Module}
AHGLM comprises the Static Adaptive Graph Learning (SAGL), the Dynamic Graph Learning (DGL) and the Spatial Gate Fusion, as illustrated in Fig.  \ref{AHSTGNN}. In the ${l^{th}}$ block, we denote the outputs of SAGL and DGL as $H_{SG}^l$ and $H_{DG}^l$. After the Spatial Gate Fusion, the output of the $l^{th}$ AHGLM is $H_S^l$.


\textbf{Static Adaptive Graph Learning.}
The Static Adaptive Graph Learning is put forward to capture the relatively stable spatial relationships between the cell towers, regardless of the geographic distance among them. We use GCN to implement the feature aggregation on graph in SAGL. According to \cite{kipf2017semi}, the graph convolution operation in the spectral domain can be well-approximated by $1^{st}$ Chebyshev polynomial expansion and generalized to high-dimensional GCN as:
\begin{equation}
    H=\left(I_{N}+D^{\frac{1}{2}} A D^{-\frac{1}{2}}\right) X \Theta,
\end{equation}
where $A$ is the adjacency matrix, $D$ is a degree matrix, $I_{N}\in \mathbb{R}^{N \times N}$ is an identity matrix, $\Theta$ is the graph convolution kernel.

We believe that the node-level attributes such as the geographic location and surrounding environment of the cell tower can be represented by the learned node embedding abstractly in the high-dimensional space. So, SAGL implements a learnable graph node embedding $E_G \in \mathbb{R}^{N \times d}$, where $d$ represents the node embedding dimension. Inspired by \cite{bai2020adaptive}, we obtain the adaptive adjacency matrix $A_{adp}$ of the static graph by multiplying $E_G$ and $E_G^T$. And the formula is as follows:
\begin{equation}
    A_{adp}=softmax(RELU(E_G\cdot E_G^T)),
\end{equation}
where $softmax$ function is used to normalize the adaptive adjacency matrix. Then, enhanced with GCN, the graph convolution operation in SAGL can be formulated as:
\begin{align}
H_{SG}^l&={(I_N+A_{adp})} H_T^{l}\Theta,
\end{align}
where the $H_{SG}^{l} \in \mathbb{R}^{T \times N \times D}$ is the output of SAGL, $\Theta\in \mathbb{R}^{D \times D}$ is the learnable graph convolution kernel.

\textbf{Dynamic Graph Learning.}
DGL implements Graph Attention Network (GAT) \cite{velivckovic2018graph} to perform dynamic spatial feature aggregation, owing to GAT's ability to capture the dynamic influences of neighbors. We believe that short-term dynamic influences mostly occur between nearby cell towers, so we deploy the exponential distance-decay matrix \cite{yu2017spatio} as the prior graph structure of GAT. We call it the distance adjacency matrix, denoted as $A_{dis} \in \mathbb{R}^{N \times N}$. As shown in Fig.  \ref{AHSTGNN}, GAT achieves weighted feature aggregation by calculating the attention scores of the central node and neighbor nodes. We also apply multi-head attention \cite{vaswani2017attention} to capture the relation semantics between nodes from different learning subspaces. Moreover, the multi-head attention can be computed in parallel for reducing time complexity. For simplicity, taking a graph node $v_i$ as an example, the graph attention mechanism of DGL can be formulated as follows:
\begin{align}
    a_{i j}&=\frac{\exp \left(L R\left(a^{T}\left[W h_{t, i} \| W h_{t, j}\right]\right)\right)}{\sum_{k \in \mathcal{N}_{i}} \exp \left(L R\left(a^{T}\left[W h_{t, i} \| W h_{t, k}\right]\right)\right)}, \\
\tilde{h}_{t, i}&=\sigma\left(\frac{1}{K} \sum_{k=1}^{K} \sum_{j \in N_{i}} a_{i j} W^{k} h_{t, j}\right),
\end{align}
where $a_{ij}\in \mathbb{R}$ represents the computed attention score between $v_i$ and $v_j$, where ${j \in \mathcal{N}_{i}}$. $\mathcal{N}_{i}$ represents the neighbors set of $v_i$ in the graph, i.e., $A_{v_i,v_j}>0$. $LR(\cdot)$ denotes the $LeakyReLU$ function. $a\in \mathbb{R}^{2D \times 1}$ is a learnable parameter vector. $W \in \mathbb{R}^{D \times D}$ is a weight matrix of a shared linear transformation. $h_{t,i} \in \mathbb{R}^D$ represents the feature of $v_i$ at time step $t$ in $H_T^l$. We use the summation to aggregate the features generated by multi-head attention. $K$ represents the number of heads and $W^{k} \in \mathbb{R}^{D \times D} $ is the transformation parameter matrix corresponding to the head $k$. $\tilde{h}_{t,i} \in \mathbb{R}^{D}$ is the DGL output of $v_i$ at time step $t$. In the $l^{th}$ block, the final output of the DGL is $H_{DG}^{l} \in \mathbb{R}^{T \times N \times D}$.

\textbf{Spatial Gate Fusion.}
We apply a Spatial Gate Fusion to fuse the two inputs $H_{SG}^l$ and $H_{DG}^l$, which come from the above mentioned SAGL and DGL in the $l^{th}$ block, respectively. As shown in Fig.  \ref{AHSTGNN}, the gate fusion mechanism can be expressed as follows:
\begin{align}
H_{S}^{l}&=gate \odot H_{S G}^{l}+(1-g a t e) \odot H_{D G}^{l}, \\
gate&=\sigma\left(H_{S G}^{l} W_{g 1}+H_{D G}^{l} W_{g 2}+b_{g}\right),
\end{align}
where $W_{g1}\in \mathbb{R}^{D \times D}$, $W_{g2}\in \mathbb{R}^{D \times D}$ and $b_g \in \mathbb{R}^{D}$ are all learnable parameters. The final output of the AHGLM is $H_{S}^l \in \mathbb{R}^{T \times N \times D}$.
The Spatial Gate Fusion can adaptively control the flow of the static and dynamic spatial dependencies at each cell tower.

\subsubsection{Spatial-Temporal Adaptive Module}
In this paper, we believe that different cell towers are affected variously by spatial and temporal influences. In other words, some cell towers may be affected more by their own data traffic than that of their neighbors. The previous spatial-temporal traffic prediction methods also noticed this problem. \cite{wu2020connecting} implement a hyper parameter $\beta$ to avoid spatial noise from neighbors, where $\beta$ is shared for all nodes. But we argue that the spatial-temporal tendencies of different nodes are different. Therefore, we propose a Spatial-Temporal Adaptive Module (STAM) to capture the node-level spatial-temporal adaptation tendencies.

\begin{figure}[t]
\begin{center}
\includegraphics[width=7cm]{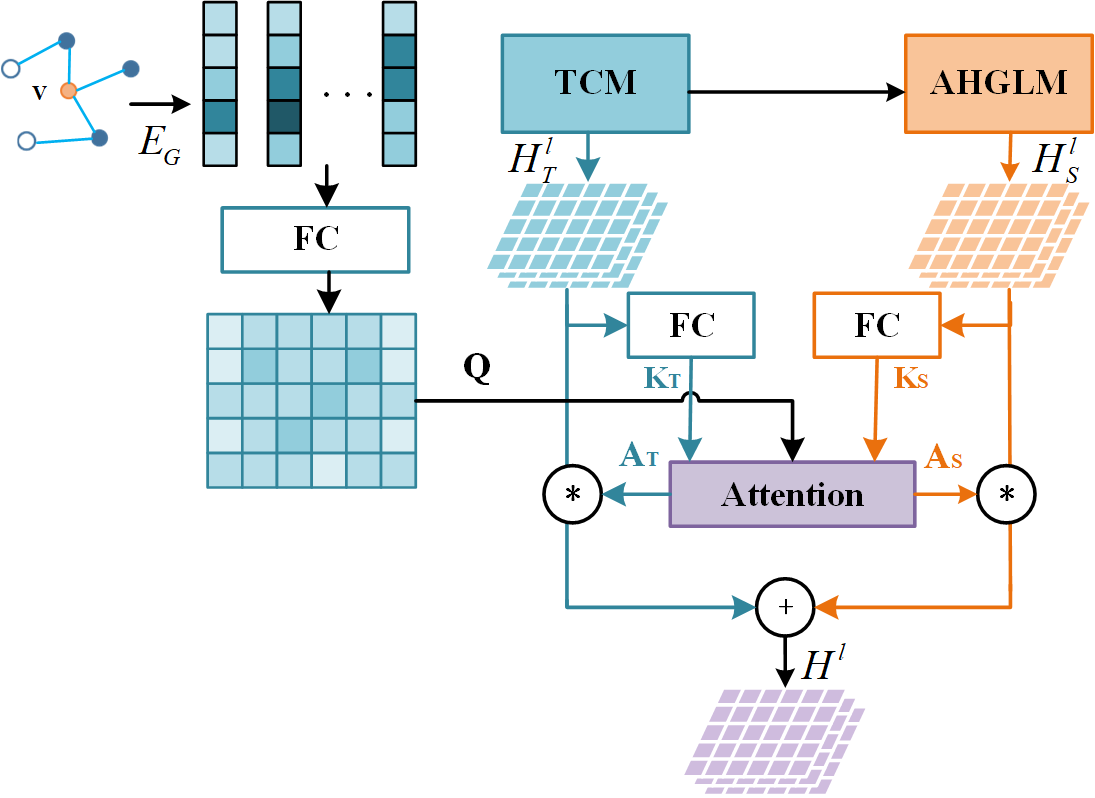}
\end{center}
\caption{Structure of Spatial-Temporal Adaptive Module.}
\label{STAM}
\end{figure}

The entire architecture of STAM is shown in Fig.  \ref{STAM}. It is worth noting that we use the same graph node embedding $E_G$ in the STAM with the AHGLM to maintain a uniform high-dimensional node representation of the cell tower in the entire model, which can also reduce the amount of parameters. First, we apply linear transformation to the node embedding $E_G$ for obtaining the graph node queries $Q=E_G W_Q$, where $W_Q\in \mathbb{R}^{d\times D}$ and $Q\in \mathbb{R}^{N\times D}$. Second, we transform the output of the TCM $H_T^{l}$ with $K_T^{l}=H_T^{l}W_{Kt}$ to get the keys of nodes in the temporal dimension, where $W_{Kt}\in \mathbb{R}^{D \times D}$. Similarly, the output of the AHGLM $H_S^l$ is transformed by $K_S^{l}=H_S^{l}W_{Ks}$ to obtain the keys of nodes in the spatial dimension, where $W_{Ks}\in \mathbb{R}^{D \times D}$. Then, we compute the attention function the same as \cite{vaswani2017attention}, formulated as $A^{l}_T = Q(K_T^l)^{T}/\sqrt{D}$ and $A^{l}_S=Q(K_S^{l})^T/\sqrt{D}$, where $A_T^{l}\in \mathbb{R}^{T\times N \times 1}$ and $A_S^{l} \in \mathbb{R}^{T\times N \times 1}$. $A^{l}_T$ and $A_{S}^l$ are the computed temporal and spatial attention of the graph nodes, respectively. At last, we calculate the output of STAM as:
\begin{align}
A t t_{T}^{l}&=\frac{\exp \left(A_{T}^{l}\right)}{\sum_{r \in\{T, S\}} \exp \left(A_{r}^{l}\right)}, \\
A t t_{S}^{l}&=\frac{\exp \left(A_{S}^{l}\right)}{\sum_{r \in\{T, S\}} \exp \left(A_{r}^{l}\right)}, \\
H^{l}&=A t t_{S}^{l} \cdot H_{S}^{l}+A t t_{T}^{l} \cdot H_{T}^{l},
\end{align}
where $Att_{T}^l \in \mathbb{R}^{T\times N \times 1}$ and $Att_{S}^l \in \mathbb{R}^{T\times N\times 1}$ are the attention scores of the nodes in the temporal and spatial dimension, respectively. $H^{l}\in \mathbb{R}^{T\times N \times D}$ is the output of STAM and is also the final output of the $l^{th}$ AHSTL block.

\subsection{Output Layer}
For each AHSTL block output, we implement skip connections to directly connect them to the output layer. With stacking multiple AHSTL blocks, the temporal receptive field of the AHSTGNN is increasing. The bottom block pays more attention to the temporally adjacent traffic features, and the high-level block focuses on long-term temporal traffic information. Skip connections are adopted to solve the spatial dependencies modeling at different temporal levels. After the skip connections, we use summation to fuse the outputs of all the AHSTL blocks, and the result is denoted as $H_{out} \in \mathbb{R}^{T\times N \times D}$. At last, we implement the output layer with two fully connected layers to generate the final multi-step prediction $\hat{Y}\in \mathbb{R}^{M \times N \times F}$. The output layer can be formulated as follows:
\begin{equation}
    \hat{Y}=RELU(H_{out}W_{f1}+b_{f1})\cdot W_{f2}+b_{f2},
\end{equation}
where $W_{f1}\in \mathbb{R}^{(TD)\times C}$, $W_{f2}\in \mathbb{R}^{C\times (MF)}$, $b_{f1} \in \mathbb{R}^C$ and $b_{f2}\in \mathbb{R}^{MF}$ are learnable parameters. $\hat{Y}$ is the final output of the entire AHSTGNN.

For the multi-step cellular traffic prediction task in this paper, we use the Mean Absolute Error (MAE) between the predicted values and the true values as the loss function and minimize it through backpropagation, formulated as:
\begin{equation}
    L(\theta)=\frac{1}{M}\sum_{i=t+1}^{t+M} |Y_i-\hat{Y}_i|,
\end{equation}
where $\theta$ represents all learnable parameters of our model. $\hat{Y}_i$ denotes the model's prediction of all cell towers at time step $i$ and $Y_i$ is the ground truth.

\section{EXPERIMENTS}



\subsection{Experiment Settings}

\begin{table}[]
 \centering
 \caption{Description of datasets.}
 \resizebox{8.5cm}{!}{
  \begin{tabular}{ccccccc}
  \hline   
 \multirow{2}{*}{Datasets} & \multirow{2}{*}{Samples} & \multirow{2}{*}{Nodes} & \multirow{2}{*}{Timespan} & \multirow{2}{*}{Timeslot} & \multirow{2}{*}{\begin{tabular}[c]{@{}c@{}}Input \\      Length\end{tabular}} & \multirow{2}{*}{\begin{tabular}[c]{@{}c@{}}Output \\      Length\end{tabular}} \\  
    &                          &                        &                           &                           &                                                                               &                                                                                \\
  \hline \text { Jiangsu } & 8640 & 1051 & \text { Jan-Mar, 2021 } & 15 min & 12 & 12 \\
  \hline \text { Milan } & 4320 & 900 & \text { Nov, 2013 } & 10 min & 6 & 6 \\
  \hline
  \end{tabular}\label{dataset}
 }
\end{table}

To evaluate the performance of our work, we conduct experiments on two real-world datasets: Jiangsu and Milan.
Specifically, we select the commonly used central regions ($30 \times 30$ grids) in Milan dataset \cite{he2020graph}.
The ratios of training set, validation set and testing set are 2:0:1 for Milan and 2:1:1 for Jiangsu. 
To facilitate comparison of the experimental results, the data split ratios, adjacency matrix, as well as the history and prediction windows of the two datasets are consistent with \cite{Wang2021adaptive}.
The detailed statistics of these two datasets are shown in Table \ref{dataset}. Our experiments are conducted on the JIUTIAN Artificial Intelligence Platform with one NVIDIA Tesla V100s GPU card using Pytorch 1.8. The hyperparameters are set as follows. For periodic input, the $L_D$ and $L_W$ are both set to 1. We stack 4 AHSTL blocks and the graph convolution kernel size is set to 2. The models are trained by the Adam optimizer. The distance adjacency matrix is constructed based on the geographic distances between nodes with thresholded Gaussian kernel, which can be formed as:
\begin{equation}
\mathbf{A}_{v_{i}, v_{j}}=\left\{\begin{array}{l}
\exp \left(-\frac{d_{v_{i}, v_{j}}^{2}}{\sigma^{2}}\right), \text { if } d_{v_{i}, v_{j}} \leq \kappa \\
0, \text { otherwise. }
\end{array}\right. 
\end{equation}
where $d_{v_{i}, v_{j}}$ denotes the distance between cell tower $i$ and $j$, $\sigma$ is the standard deviation and $\kappa$ is the threshold used to control the sparsity of the adjacency matrix.


We compare AHSTGNN \footnote{Code available at: https://github.com/starxingwang/AHSTGNN} with typical cellular traffic prediction methods including HA (Historical Average), LSTM\cite{6795963}, MVSTGN\cite{9625773} and AMF-STGCN\cite{Wang2021adaptive}, as well as the generic advanced spatial-temporal sequence prediction methods including Graph Wavenet\cite{ijcai2019-264}, MTGNN\cite{wu2020connecting} and AGCRN\cite{bai2020adaptive}.
The Mean Absolute Error (MAE) and Root Mean Square Error (RMSE) are used to evaluate the performance of different models.

\subsection{Experiment Results}

 

\begin{table}[htbp]
\caption{Performance comparsion of AHSTGNN and baseline models on Jiangsu and Milan datasets.}
\setlength{\tabcolsep}{4mm}{
 \centering
 
 \begin{tabular}{ccccc}
  \hline
  \multirow{2}[2]{*}{Models} & \multicolumn{2}{c}{Jiangsu} & \multicolumn{2}{c}{Milan} \\
  \cmidrule{2-5}   &  MAE & RMSE   & MAE   & RMSE    \\
  \midrule
    HA &196.76  & 334.74 &61.28 & 120.73 \\
    LSTM & 168.46 & 313.21  & 43.28  & 79.77  \\
    Graph WaveNet & 130.83 & 254.28 & 32.71 & 65.35  \\
    MVSTGN & 164.14 & 319.57& 35.03 & 70.29  \\
    MTGNN &   130.61    & 253.26  &   \underline{29.13}   &    \underline{57.63}   \\
    AGCRN & \underline{129.01} & \underline{251.19} & 30.27 & 59.81   \\
    AMF-STGCN & 129.78 & 252.71 & 30.59 & 57.89  \\
    AHSTGNN  & \textbf{124.81} & \textbf{243.91}& \textbf{27.96} & \textbf{57.18}  \\
  \bottomrule
 \end{tabular} \label{basejm}
}
\end{table}

Table \ref{basejm} shows the overall performance on Jiangsu and Milan datasets. Our AHSTGNN achieves the best performance on both datasets.
We observe that HA has the worst prediction performance since it can hardly capture the nonlinear temporal dependencies in cellular traffic. LSTM outperforms HA but is inferior to other spatial-temporal prediction models.
The GCN-based methods such as Graph WaveNet, MTGNN and AGCRN, show impressive performance on spatial-temporal cellular traffic prediction.
Although Graph WaveNet and AGCRN present to learn an adaptive graph adjacency matrix, the adaptive graphs that they learn are static and can barely reflect the time-varying spatial correlations among cell towers and they perform slightly worse than AHSTGNN. AHTSTGNN considers both the stable relationships and dynamic influences among nodes, and is more capable of learning the complicated spatial dependencies in cellular traffic data.

\begin{figure}[htpb]
 \centering
 \includegraphics[width=9cm]{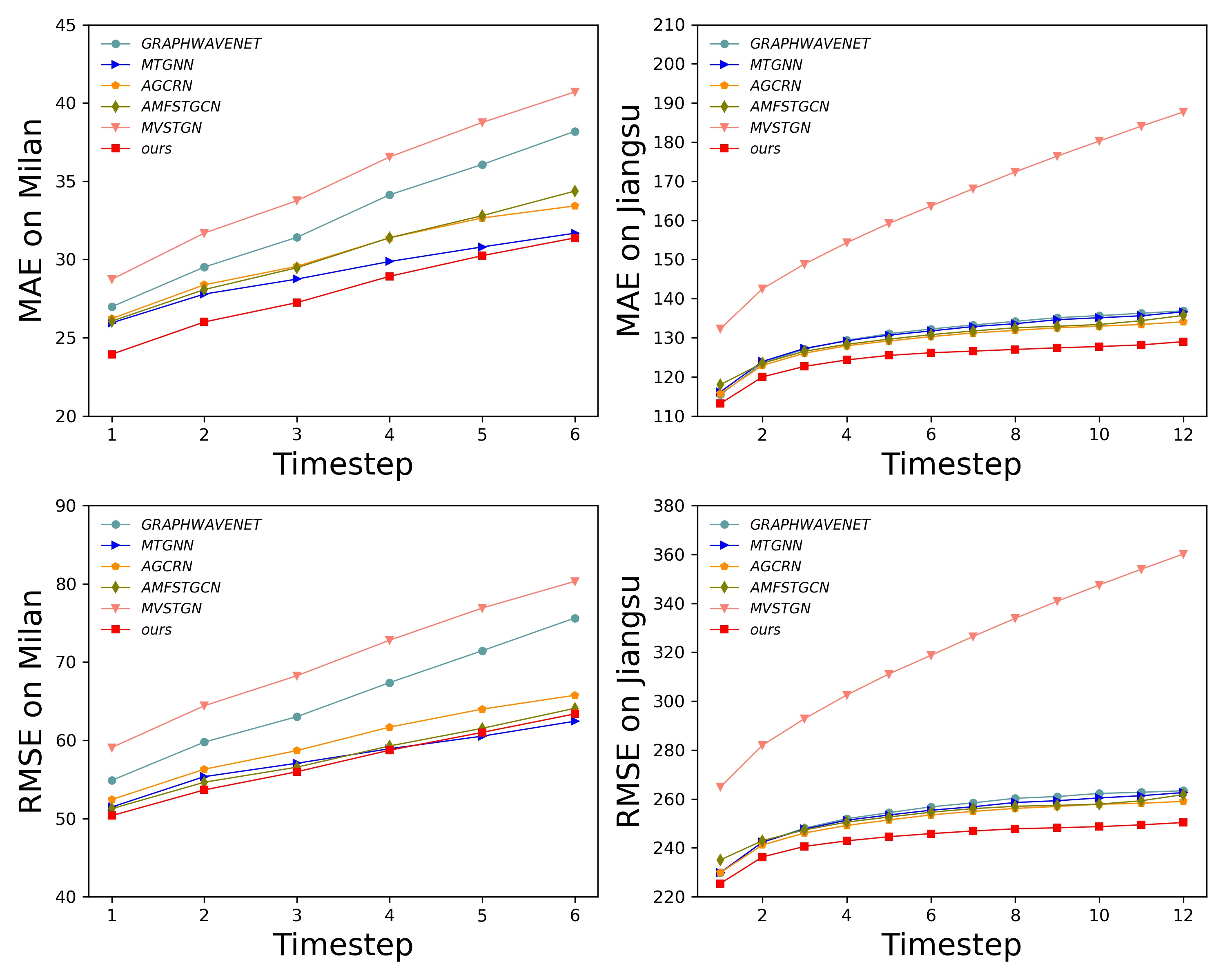}
 \caption{Prediction performance comparison at each horizon on Jiangsu and Milan datasets.}
 \label{result}
\end{figure}

The results of MVSTGN are not ideal due to the fact that it relies heavily on the attention mechanism and is therefore susceptible to data noise and volatile data. Since the Jiangsu dataset is more dynamic than the Milan dataset, the corresponding performance of MVSTGN on Jiangsu dataset is even worse. AMF-STGCN shows relatively better performance on both datasets, indicating that the spatial-temporal correlations and heterogeneity modeling can bring benefits for cellular traffic prediction.
Fig. \ref{result} depicts the prediction performance at each time step for both datasets. AHSTGNN achieves the best performance in both short-term and long-term prediction, and maintains a relatively stable error even at the last time step. The significant increase in performances on the Jiangsu dataset also indicates the effectiveness of AHSTGNN in handling cellular traffic data with complex spatial-temporal dependencies.

\subsection{Ablation Study}
To further evaluate the effectiveness of different modules in AHSTGNN. We conduct ablation experiments with four variants of AHSTGNN:
\begin{itemize}
\item DGL-adp+dis w/o SAGL: We deploy the DGL on adaptive adjacency matrix and distance adjacency matrix, and remove the SAGL of AHGLM from AHSTGNN.
\item SAGL-adp+dis w/o DGL: We deploy the SAGL on adaptive adjacency matrix and distance adjacency matrix, and remove the DGL of AHGLM from AHSTGNN.
\item w/o STAM: This is AHSTGNN whose node-level Spatial-Temporal Adaptive Module is removed.
\item w/o TCM-periodic data: We replace the three periodic data inputs with recently data traffic input only.
\end{itemize}
\par

\begin{table}[htbp]
	\caption{Ablation study.}
		\centering
		\begin{tabular}{ccccc}
			\toprule
			\multirow{2}[4]{*}{Model} & \multicolumn{2}{c}{Jiangsu} & \multicolumn{2}{c}{Milan} \\
			\cmidrule{2-5}  &        
            \multicolumn{1}{c}{MAE} &  \multicolumn{1}{c}{RMSE} & \multicolumn{1}{c}{MAE} &  \multicolumn{1}{c}{RMSE}   \\
			\midrule
			AHSTGNN  &  \textbf{124.81}       &   \textbf{243.91}    &   \textbf{28.57}       & \textbf{57.91}       \\
			{DGL-adp+dis w/o SAGL} & \underline{124.84}         &   \underline{246.61}    &   29.09   & \underline{58.23}  \\
            {SAGL-adp+dis w/o DGL} &   125.42      &  248.28     &   29.35  & 59.08   \\
            {w/o STAM} &   127.97       &  250.06     &    32.80   & 63.04  \\
            {w/o TCM-periodic data} &   132.09       &  256.86     &   \underline{28.67}     &  59.05    \\

			\bottomrule  
		\end{tabular}\label{ablation}
\end{table}


As shown in the Table \ref{ablation}, the experimental results on Jiangsu and Milan datasets are illustrated,  and we observe that: 
1) The results for both SAGL-adp+dis w/o DGL and DGL-adp+dis w/o SAGL are worse than AHSTGNN, indicating the effectiveness of the DGL and SAGL. 
DGL can capture the irregular dynamic spatial correlations and SAGL is able to exploit the stable interactions between the cell towers. 
2) STAM module can effectively solve the problem of node heterogeneity and improve the model performance, as the module effectively protecting the nodes from unexpected spatial noise.
3) The periodic data input brings more benefits than recently data traffic only, as the cellular traffic datasets inherit strong daily and weekly periodic characteristics.
In summary, each component of our AHSTGNN has benefits for modeling the crucial aspects of traffic data, and they jointly promote the cellular traffic prediction performance.

\subsection{Computational Complexity}

\begin{table}[htbp]
\caption{The computation cost on the Milan dataset.}
\resizebox{8.5cm}{!}{
\centering
\begin{tabular}{ccc}
 \hline Model & Parameters & Training Time (s/epoch) \\
 \hline 
 AHSTGNN & 255862 & 10.46  \\
 Graph WaveNet & 236438 & 9.99 \\
 MVSTGN & 493401 & 35.2  \\
 AGCRN & 754350  & 10.02 \\
AMF-STGCN & 982393 & 36.53\\
\hline
\end{tabular}\label{time}
}
\end{table}

To evaluate the computational complexity, we compare the parameter numbers and training time of AHSTGNN with Graph WaveNet, MVSTGN, AGCRN and AMF-STGCN on the Milan dataset in Table \ref{time}. Due to the fact that MTGNN has been optimized in the model training process, it is unfair to compare it to other algorithms, so we do not include it. As shown in Table \ref{time}, AHSTGNN requires a slightly higher number of parameters than Graph WaveNet as a tradeoff for hybrid graph learning. 
AHSTGNN is slightly slower than Graph WaveNet and AGCRN in terms of training time, as a sacrifice for improving the prediction performance. Graph WaveNet and AGCRN maintain relatively simple model structures, consisting only of the graph convolutions and temporal convolutions or the recurrent networks.
MVSTGN and AMF-STGCN take longer to train than other models, more than three times that of AHSTGNN. We hypothesize that one of the primary reasons is that they employ the complicated model architectures to learn the intricate spatial-temporal correlations.
In conclusion, the computation cost of AHSTGNN is reasonable in light of the improvement in accuracy of prediction that AHSTGNN provides.

\section{CONCLUSION}
In this paper, we propose a novel AHSTGNN for cellular traffic prediction. We present an AHGLM which combines the static adaptive graph learning with dynamic graph learning to capture the compound spatial dependencies. We employ the TCM with multi-periodic data inputs to capture the non-linear temporal dependencies of cellular data traffic. Besides, a spatial-temporal adaptive module is implemented to tackle the heterogeneity problem. Our experiments on two real-world cellular traffic datasets show that AHSTGNN achieves the state-of-the-art, demonstrating the superior traffic prediction capabilities of our proposed model. In the future, we plan to go deeper into cellular traffic prediction by exploring the impact of crowds' migration on cellular traffic.

\end{document}